\documentclass[sigplan,10pt,nonacm]{acmart}

\usepackage[]{hyperref}
\usepackage{graphicx}
\usepackage{algorithm}
\usepackage{algpseudocode}
\usepackage{soul}
\usepackage{color}
\usepackage{enumitem}
\usepackage{algorithm}
\usepackage{algpseudocode}
\usepackage{multirow}
\usepackage{xcolor} 
\usepackage{hyperref} 
\hypersetup{hidelinks} 

\soulregister{\cite}{7} 
\soulregister{\ref}{7}  
\soulregister{\pageref}{7} 

\algnewcommand{\LeftComment}[1]{\Statex \(\triangleright\) #1}

\algnewcommand\algorithmicforeach{\textbf{for each}}
\algdef{S}[FOR]{ForEach}[1]{\algorithmicforeach\ #1\ \algorithmicdo}

\acmConference[PL'18]{ACM SIGPLAN Conference on Programming Languages}{January 01--03, 2018}{New York, NY, USA}
\acmYear{2018}
\acmISBN{} 
\acmDOI{} 
\startPage{1}
\setcopyright{none}
\usepackage{booktabs}   
\usepackage{subcaption} 

\algblock{ParFor}{EndParFor}
\algnewcommand\algorithmicparfor{\textbf{parfor}}
\algnewcommand\algorithmicpardo{\textbf{do}}
\algnewcommand\algorithmicendparfor{\textbf{end\ parfor}}
\algrenewtext{ParFor}[1]{\algorithmicparfor\ #1\ \algorithmicpardo}
\algrenewtext{EndParFor}{\algorithmicendparfor}

\begin{document}

\title[]{Analysis of Stable Vertex Values: \\ Fast Query Evaluation Over An Evolving Graph} 


\author{Mahbod Afarin}
\authornote{Both authors contributed equally to this research.}
\email{mafar001@ucr.edu}
\affiliation{%
  \institution{CSE Department, UC Riverside}
  \country{USA}
}

\author{Chao Gao}
\authornotemark[1]
\email{cgao037@ucr.edu}
\affiliation{%
  \institution{CSE Department, UC Riverside}
  \country{USA}
}

\author{Xizhe Yin}
\email{xyin014@ucr.edu}
\affiliation{%
  \institution{CSE Department, UC Riverside}
  \country{USA}
}

\author{Zhijia Zhao}
\email{zhijia@cs.ucr.edu}
\affiliation{%
  \institution{CSE Department, UC Riverside}
  \country{USA}
}

\author{Nael Abu-Ghazaleh}
\email{nael@cs.ucr.edu}
\affiliation{%
  \institution{CSE Department, UC Riverside}
  \country{USA}
}

\author{Rajiv Gupta}
\email{rajivg@ucr.edu}
\affiliation{%
  \institution{CSE Department, UC Riverside}
  \country{USA}
}

\begin{abstract}
Evaluating a query over a large, irregular graph is inherently challenging. This challenge intensifies when solving a query over a sequence of snapshots of an evolving graph, where changes occur through the addition and deletion of edges. We carried out a study that shows that due to the gradually changing nature of evolving graphs, when a vertex-specific query (e.g., SSSP) is evaluated over a sequence of 25 to 100 snapshots, for 53.2\% to 99.8\% of vertices, the query results remain unchanged across all snapshots. Therefore, the \emph{Unchanged Vertex Values} (UVVs) can be computed once and then minimal analysis can be performed for each snapshot to obtain the results for the remaining vertices in that snapshot. We develop a novel \emph{intersection-union analysis} that very accurately computes lower and upper bounds of vertex values across all snapshots. When the lower and upper bounds for a vertex are found to be equal, we can safely conclude that the value found for the vertex remains the same across all snapshots. Therefore, the rest of our query evaluation is limited to computing values across snapshots for vertices whose bounds do not match.  We optimize this latter step evaluation by concurrently performing incremental computations on all snapshots over a significantly smaller subgraph. Our experiments with several benchmarks and graphs show that we need to carry out per snapshot incremental analysis for under 42\% vertices on a graph with under 32\% of edges. Our approach delivers speedups of 2.01-12.23$\times$ when compared to the state-of-the-art RisGraph implementation of the KickStarter-based incremental algorithm for 64 snapshots.

\end{abstract}




\maketitle 

\section{Introduction}

Graph analytics are employed in many domains to uncover insights from connected data. There has been much work resulting in scalable graph analytics systems for GPUs, multicore servers, and  clusters~\cite{pregel, graphlab, powergraph, galois, ligra, coral, gridgraph, graphchi, cusha, cusha2, tigr, gunrock, pingali, saday}. Most real-world graphs change dynamically over time~\cite{sakr-21}. Therefore, recently there has been a great deal of interest in analytics over changing graphs~\cite{kickstarter,graphbolt,risgraph,CommonGraph,tegra,evog-taco,chronos}. Efficient dynamic graph processing has diverse applications, including social network analysis for community detection and influence propagation~\cite{aggarwal2014, giatsidis2011d}, personalized recommendation systems~\cite{amatriain2015recommender, koren2009matrix}, and telecommunication networks for traffic management and fault detection~\cite{nguyen2018survey, li2017survey}. It is also crucial in financial networks for fraud detection and risk assessment~\cite{akoglu2015graph, kumar2016edge}, biological networks like gene regulatory networks~\cite{liu2018dynamic, zhang2012inferring}, and transportation networks for traffic flow optimization~\cite{wang2016dynamic, zheng2013u}. In e-commerce, it aids in customer interaction analysis and supply chain management~\cite{yan2017survey, ngai2012information}, while in cybersecurity, it enhances intrusion detection and network defense~\cite{ahmed2016survey, garcia2014empirical}. Additionally, smart cities benefit from urban planning and resource management applications~\cite{batty2012smart, neirotti2014current}, and healthcare uses include epidemic tracking and patient monitoring~\cite{he2016dynamic, rumsfeld2016big}. These applications underscore the importance of dynamic graph processing.


As a graph continues to evolve, a sequence of snapshots is captured which grows in length over time. An \emph{evolving graph query} is aimed at analyzing the evolution of a graph property (e.g., shortest paths) over a time window. That is, an evolving graph query typically requires a graph query to be solved over a sequence of snapshots $G_0, G_{1}, \ldots, G_{n}$. By selecting the time window, the user requests query evaluation for all snapshots within a time window to observe trends in query results (e.g., changes in shortest paths). As the duration of the time window for query evaluation increases, so do the number of snapshots that must be analyzed and hence the cost of query evaluation rises. Thus, efficiently evaluating a query over many snapshots is an important open problem.

\vspace{0.05in}
\noindent
\textbf{Existing Approaches.~} To reduce the high cost of query evaluation over a large number of snapshots, existing approaches such as Tegra~\cite{tegra} and CommonGraph~\cite{CommonGraph} leverage incremental algorithms. A general incremental algorithm that supports both edge additions and deletions was first proposed in KickStarter~\cite{kickstarter} and then further extended and optimized by Graphbolt~\cite{graphbolt} and RisGraph~\cite{risgraph} respectively. While both Tegra and CommonGraph employ incremental algorithms, there is a major difference. Tegra explicitly processes both additions and deletions using incremental algorithms developed in KickStarter~\cite{kickstarter} and Graphbolt~\cite{graphbolt}. In contrast, CommonGraph~\cite{CommonGraph} converts edge deletions between snapshots into edge additions between a common graph and each snapshot, thus trading expensive deletions processing for relatively inexpensive additions processing. The common graph is first used to evaluate a query. Starting from the common graph and query results computed for it, edge additions are processed to incrementally compute query results for each snapshot. 



\vspace{0.05in}
\noindent
\textbf{Our Insight: UVVs.~} We develop a new insight and a novel algorithm to take advantage of the insight to substantially optimize query evaluation across multiple snapshots. 

\begin{figure}[!t]
    \centering
    \includegraphics[width=0.9\linewidth]{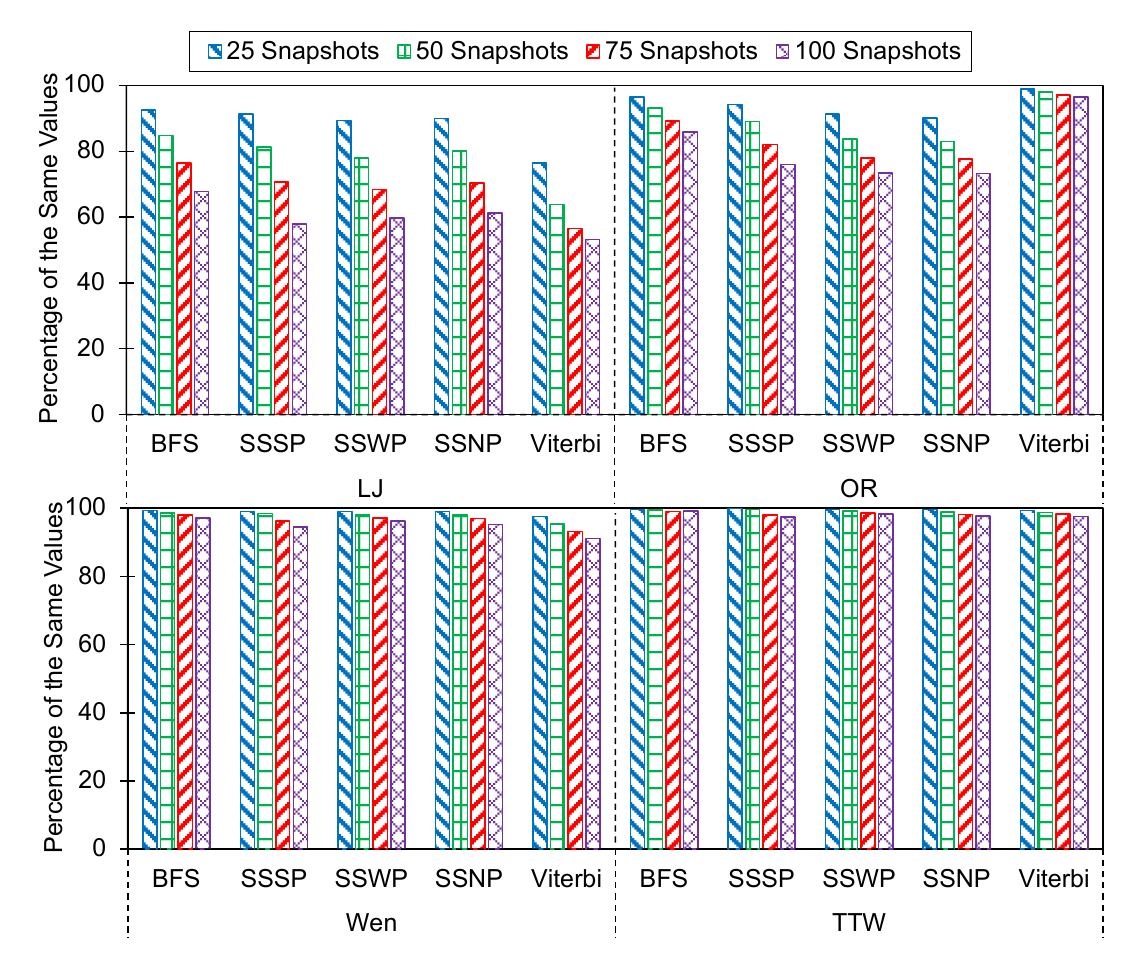}
    \vspace{-0.175in}
    \caption{Given series of 100 snapshots obtained by performing 100K edge updates (50\% deletions and 50\% additions) from one snapshot to the next, the above plot gives the percentage of vertex property values that remain unchanged across 25, 50, 75, and 100 snapshots for 4 input graphs and 5 benchmarks. 100K edges represent between 0.14\% of edges in LJ to 0.025\% for Wen.}
    \label{motivation1}
    \vspace{-0.175in}
\end{figure}

Our key insight originates from the gradually changing nature of an evolving graph. From the results of a study presented in Figure~\ref{motivation1}, we observe that, \textbf{given a vertex-specific query (e.g., SSSP), the query results for 53.2\% to 99.8\% of vertices remain unchanged across  25 to 100 consecutive snapshots.} Therefore the \emph{unchanged vertex values} (UVVs) can be computed once, and then minimal analysis can be performed for each snapshot to obtain the results for the remaining vertices in that snapshot. Moreover, the incremental computation can be performed over a much \emph{smaller graph} obtained by eliminating the incoming edges of all vertices with unchanged values -- since no updates need to be applied to their values, the incoming edges play no role in the incremental computation for any snapshot.

\vspace{0.05in}
\noindent
\textbf{Our Solution.~}
To take advantage of the above insight, and resulting opportunities for optimizing evaluation of an evolving graph query, we need to address the following key challenge. Given a vertex-specific query $Q$ and a sequence of snapshots, we must \textbf{identify vertices with unchanged vertex values} (UVVs). To address this challenge, we develop a novel \textbf{intersection-union analysis} to compute lower and upper bounds of a vertex value across all snapshots. When the bounds are found to be equal, we can \textbf{safely} conclude that the vertex value found remains the same across all the provided snapshots. Therefore, the rest of our query evaluation is limited to computing values across snapshots for vertices whose bounds were not equal. Our approach is applicable to path-based monotonic algorithms. A path-based monotonic algorithm incrementally explores or constructs paths within a graph while ensuring that the computed metric along each path (e.g., distance, cost) adheres to a monotonic property, such as non-increasing or non-decreasing values. As the algorithm progresses, it extends partial paths in such a way that the solution's quality improves or remains constant, ensuring convergence to an optimal or sub-optimal solution without regressing to a worse state.

We further optimize evaluation of query $Q$ by \emph{concurrently} performing incremental computations for all snapshots over a significantly smaller graph that we refer to as the $Q$-Relevant SubGraph ($QRS$). The smaller graph is obtained by eliminating the incoming edges of all vertices with unchanged values.

Our experiments with several benchmarks and graphs for 64 snapshots show that we need to carry out per snapshot incremental analysis for under 42\% vertices on a graph with under 32\% of edges. When we incrementally evaluate the query on each snapshot using $QRS$, the cost of evaluation is lowered. 
Our approach delivers speedups of up to 12.23$\times$ over the state-of-the-art RisGraph implementation of the KickStarter-based incremental algorithm for 64 snapshots.

\vspace{0.05in}
\noindent
The key contributions of this paper are as follows:
\begin{itemize}[leftmargin=*, topsep=1pt]
\setlength{\itemsep}{0.5pt}
\setlength{\parskip}{0.5pt}
    \item
    \textsf{\it Identifying Unchanged Vertex Values:} We 
    develop a novel \emph{intersection-union analysis} for identifying unchanged vertex values by bounding vertex values across all snapshots. 
    
    \item
    \textsf{\it Reduced Graph for Incremental Computation:} We identify a significantly smaller $Q$-Relevant Subgraph ($QRS$) over which incremental computations are performed.
    
    \item
    \textsf{\it Concurrent Incremental Evaluation for All Snapshots:} We build a system that simultaneously performs incremental computations for all snapshots further reducing the cost. 

    \item 
    \textsf{\it Experimental Evaluation:} We demonstrate our approach by applying it to evaluation of queries across 64 snapshots of five input graphs and five monotonic algorithms.  
\end{itemize}
\section{Background}

An evolving graph consists of a series of snapshots $G_0$, $G_1$ $\cdots$ $G_n$ of a graph captured over time. 
In evolving graph analytics, we are interested in solving a query over a specific time window during which the graph is evolving. Multiple snapshots of the graph at different points in time during the time window are available, and solving a query over a time window requires computing its results for all the snapshots within that window. Evolving graph analytics is motivated by a user's need to observe trends in a graph property. A naive approach for evaluating a query over all snapshots, shown in Figure~\ref{fig:evolving_approaches}(a), simply evaluates a query from scratch on each snapshot. To overcome  the obvious inefficiency of this approach, the following \emph{incremental approaches} have been proposed: the KickStarter-based  streaming approach; and the Common Graph deletion-free approach.

\subsection{KickStarter-based Incremental Approach}

\begin{figure}[!h]
    \centering
    \includegraphics[width=0.8\columnwidth]{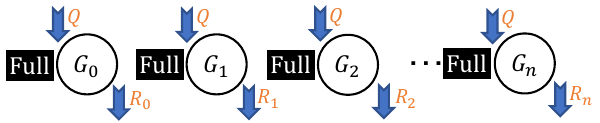}

    \vspace{0.05in}
    (a) \textsf{Naive Technique}: Full Computation (\textbf{Full}) on each snapshot of the graph ($G_0$, $G_1$, ..., $G_n$) and independently calculate the results for each snapshot from scratch.
    \vspace{0.15in}
    
    \includegraphics[width=0.7\columnwidth]{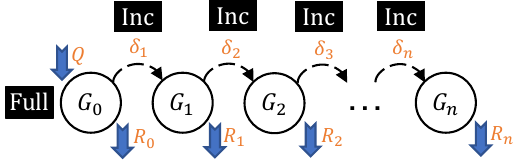}

    \vspace{0.05in}
    (b) \textsf{Kickstarter-based Incremental}: Full Computation (\textbf{Full}) on the first snapshot of the graph ($G_0$), and Incrementally (\textbf{Inc}) apply the delta batches ($\delta_1$, $\delta_2$, ..., $\delta_n$) to find the results for each snapshot in ($G_0 \rightarrow G_1 \rightarrow \cdots G_n$).
    \vspace{0.1in}

    \includegraphics[width=0.6\columnwidth]{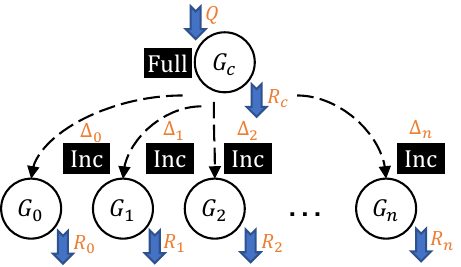}

    \vspace{0.05in}
    (c) \textsf{Deletion-Free Common Graph}: Full Computation (\textbf{Full}) on the Common Graph and Incrementally (\textbf{Inc}) add the delta batches ($\Delta_0$, $\Delta_1$, ..., $\Delta_n$) to the Common Graph to find the results for each snapshot of the graph ($G_0$, $G_1$, ..., $G_n$).
    \vspace{-0.05in}
    
    \caption{Strategies for Evolving Graph Query Evaluation.}
    \label{fig:evolving_approaches}
    \vspace{-0.25in}
\end{figure}

Without loss in generality, we assume that all vertices are present in all snapshots and the changes from one snapshot to the next are represented in the form of additions and deletions of the edges applied to an earlier snapshot that produces the next snapshot. The batches of edges, including additions and deletions, are denoted as $\delta_1, \delta_2 \cdots \delta_n$ in Figure~\ref{fig:evolving_approaches}(b). The KickStarter-based~\cite{kickstarter} incremental approach evaluates the query of the first snapshot $G_0$ from scratch and then incrementally processes $\delta_1$ through $\delta_n$ in turn to obtain query results for $G_1$ through $G_n$ as shown in Figure~\ref{fig:evolving_approaches}(b).

\subsection{The Common Graph Deletion-Free Approach}
While the KickStarter-based approach avoids redundant computation in the naive approach, it still needs to pay the high cost of evaluation associated with processing edge deletions. Past work on JetStream~\cite{jetstream} has shown that incremental processing for an edge deletion operation is significantly more expensive than the edge addition operation for monotonic graph queries. Therefore, the Common Graph approach was recently proposed to avoid both redundant computation and expensive handling of deletions (see Figure~\ref{fig:evolving_approaches}(c)). Common Graph is the subgraph that is shared by all the snapshots under consideration. Therefore, solving the query on it, and then streaming different batches of edge additions enables incrementally computing the query on each snapshot without having to explicitly deal with edge deletions. 

\emph{Common Graph} is the subgraph that is common to all snapshots of the evolving graph. Therefore, each snapshot can be obtained by simply adding an appropriate subset of edges to the Common Graph, that is, no edge deletions are required to convert the Common Graph to any snapshot. After computing the query on this Common Graph, by adding the missing edges for a snapshot and incrementally updating the query results in response to the additions, the query results for the snapshot are obtained. This is called the \emph{direct hop} approach. 

\begin{figure}[!t]
    \centering
    \includegraphics[width=0.9\linewidth]{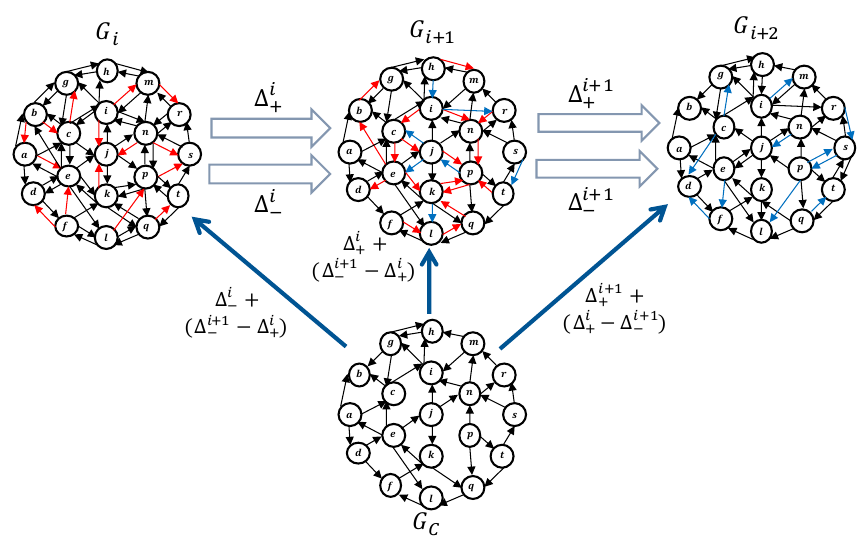}
    \vspace{-0.15in}
    \caption{Common Graph $G_{C}$ of snapshots $G_i$, $G_{i+1}$, $G_{i+2}$.}
    \label{fig:commongraph}
    \vspace{-0.1in}
\end{figure}

Figure~\ref{fig:commongraph} shows the Common Graph $G_{C}$ for three snapshots $G_i$, $G_{i+1}$, and $G_{i+2}$. We add $\Delta_{+}^{i}$ and remove $\Delta_{-}^{i}$ edges to incrementally derive $G_{i+1}$ from $G_i$. Similarly, we can derive $G_{i+2}$ from $G_{i+1}$ by respectively adding and deleting the delta batches of edges. The \emph{Common Graph} for the three snapshots, $G_{C}$, is also shown. \emph{Direct hop} approach adds different subsets of edges to the Common Graph to derive the three snapshots as shown in the figure.  For example, to derive $G_i$ we should combine three batches of edges ($\Delta_{-}^{i}$, $\Delta_{-}^{i+1}$, and $\Delta_{-}^{i+2}$) and apply them once to $G_{C}$. The main strength of the \emph{direct hop} workflow is that we can add all the addition delta batches independently and find all the snapshots in parallel. The main limitation of the \emph{direct hop} is the redundant addition operations. For example, to derive $G_i$ and $G_{i+1}$ we must add $\Delta_{-}^{i+1}$ and $\Delta_{-}^{i+2}$ twice to $G_C$. Therefore \emph{work sharing} was proposed to further reduce redundant additions. 

Though Common Graph provides significant speedups over KickStarter-based method~\cite{CommonGraph}, its scalability is limited. 
Therefore, we argue that to further optimize performance, it is essential to \textbf{eliminate wasteful work on analyzing UVV vertices and traversing incoming edges of UVV vertices in the Common Graph}. In subsequent sections, we demonstrate how to identify UVVs and exploit them to improve scalability.

\section{Our Approach Based Upon Identifying Unchanged Vertex Values (UVVs)}

\begin{figure}[!t]
    \centering
    \includegraphics[width=0.99\linewidth]{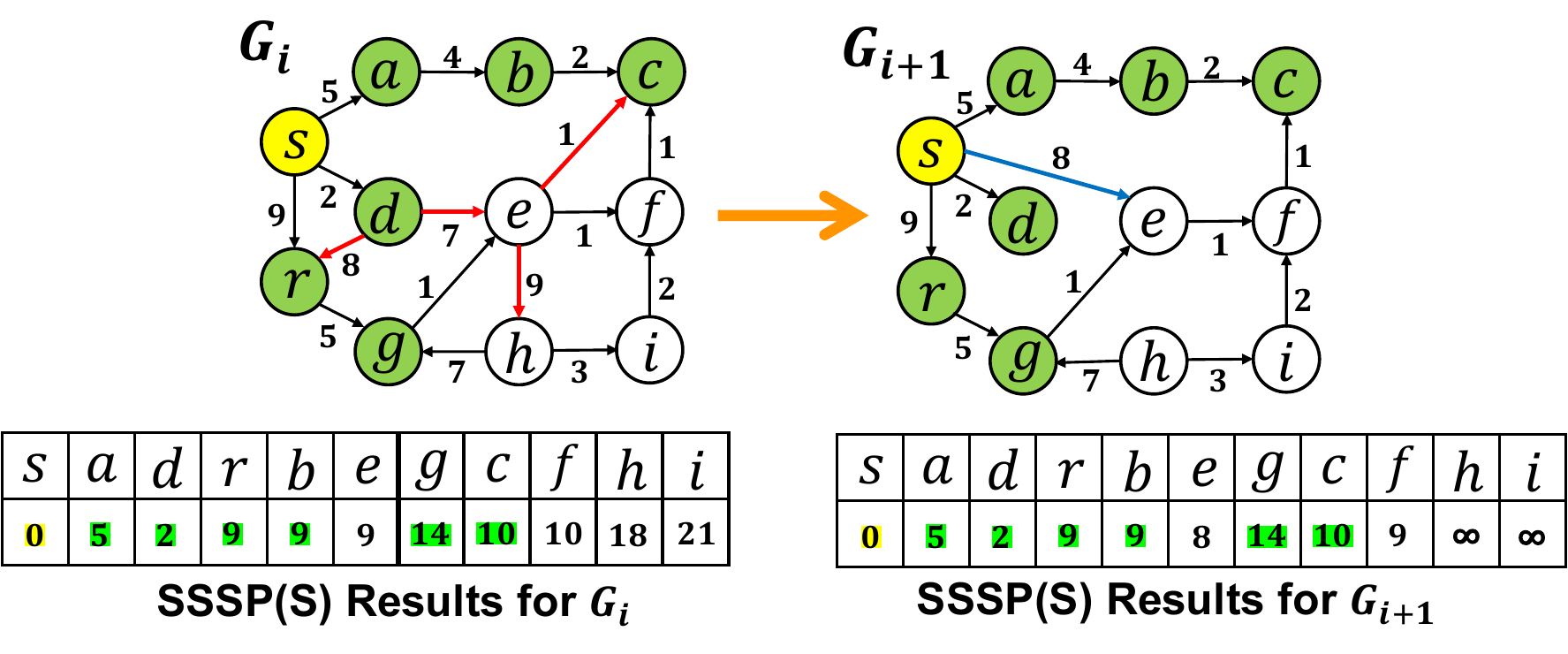}
    \vspace{-0.1in}
    \caption{Results of query SSSP(s) for two consecutive snapshots. The second snapshot is obtained by deleting red edges from the first snapshot and also adding the blue edges to the second snapshot. Note that the shortest path values for vertices marked in green are identical for both snapshots.}
    \label{alg1}
    \vspace{-0.175in}
\end{figure}

From our motivating study, we observed that the query results computed for different snapshots are substantially the same, i.e., the addition and deletion of edges frequently causes changes to property values of a small subset of vertices. In our discussion, we use UVVs to refer to vertices with Unchanged Vertex Values. The example in Figure~\ref{alg1} illustrates the presence of UVVs -- 6 of the 10 total shortest path values computed from source vertex $s$ are the same for the two snapshots, these are the ones marked green. 

\vspace{0.125in}
\begin{tabular} {|p{7.5cm}|} \hline
\textsf{Put differently, we observe that, for a specific query $Q$, there are many vertices whose property values remain unchanged across all snapshots.} \\ \hline
\end{tabular}
\vspace{0.15in}

This observation motivates us to identify UVVs and then use them to eliminate wasteful work performed during incremental computations, including computations that attempt to update UVV vertices and edge traversals that lead to UVV vertices. By identifying UVVs and shrinking the size of the graph over which incremental computations for a given query $Q$ are performed, we will affect this optimization. The reduced graph is named Q-Relevant Subgraph (QRS) for query $Q$. Next we describe the steps of our algorithm and illustrate it using an example.

Let us consider the shortest path query in this discussion, though our approach is applicable to other vertex specific path-based monotonic algorithms. Furthermore, without loss of generality, assume that all vertices are present in all snapshots. Only the set of edges present differs across the snapshots due to batches of updates in the form of edge additions and deletions that are performed as the graph evolves.

\vspace{0.1in}
Our approach for identifying UVVs and generating the query specific Q-Relevant Subgraph is as follows.

\paragraph{Step 1: Bounding Vertex Values for a Query} 
Let $E_0$, $E_1$, $\cdots$ $E_n$ denote the sets of edges corresponding to the evolving graph's snapshots $G_0$, $G_1$, $\cdots$ $G_n$. We consider two graphs that are derived from the above snapshots as follows:

\begin{itemize}\itemsep2pt
    \item \textbf{Intersection Graph} $G_\cap$: This is the graph that contains edges that are common to all the snapshots, i.e., $E_\cap = E_0 \cap \cdots \cap E_n$.
    \item \textbf{Union Graph} $G_\cup$: This is the graph that contains all edges present across all the snapshots, i.e., $E_\cup = E_0 \cup \cdots \cup E_n$.
\end{itemize}

\begin{figure}[!h]
    \centering
    \includegraphics[width=0.99\linewidth]{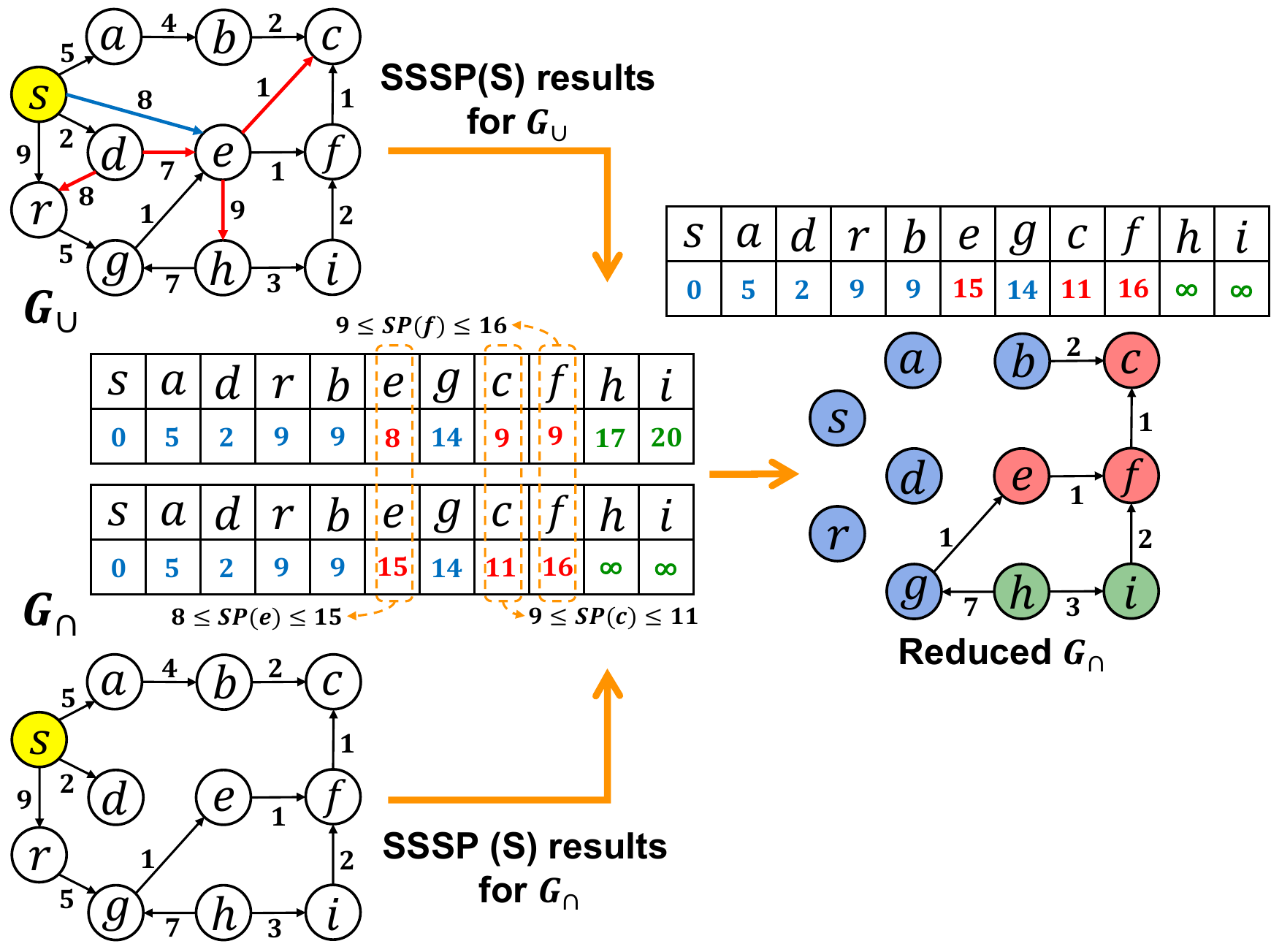}
    \vspace{-0.1in}
    \caption{The Union Graph $G_{\cup}$ provides upperbounds on path lengths across all snapshots while the Intersection Graph $G_{\cap}$ provides the lowerbounds. Query Relevant Graph obtained by reducing $G_{\cap}$ and the query results used to bootstrap incremental computations.}
    \label{alg2}
    \vspace{-0.15in}
\end{figure}

We evaluate the shortest path query for source vertex $s$ on both $G_\cap$ and $G_\cup$. Let us denote the shortest path value computed for some vertex $v$ corresponding to  $G_\cap$ and $G_\cup$ by $Val_\cap(s,v)$ and $Val_\cup(s,v)$. The following theorem captures the relationship between the shortest path value of $v$ for any snapshot $G_i$.

\vspace{0.1in}
\textbf{Theorem 1.} Given source vertex $s$, the shortest path values from $s$ to vertex $v$ for $G_\cap$ and $G_\cup$, that is,  $Val_\cap(s,v)$ and $Val_\cup(s,v)$, represent the \emph{upperbound} and \emph{lowerbound} over the shortest path value of vertex $v$ across all snapshots $G_0$, $G_1$ $\cdots$ $G_n$.

\textbf{Proof.} 
We observe that the Intersection Graph $G_\cap$ contains only a subset of paths from any snapshot $G_i$ because $E_\cap = E_0 \cap \cdots \cap E_n$. Therefore, the shortest path value of vertex $v$ corresponding to snapshot $G_i$, denoted by $Val_i(s,v)$, is bounded by $Val_\cap(s,v)$ as follows:

\vspace{-0.15in}
\[  Val_i(s,v) \leq Val_\cap(s,v) \]

Similarly, we observe that the Union Graph $G_\cup$ contains a superset of paths from any snapshot $G_i$ because $E_\cup = E_0 \cup \cdots \cup E_n$. Therefore, the shortest path value of vertex $v$ corresponding to snapshot $G_i$, denoted by $Val_i(s,v)$, is bounded $Val_\cup(s,v)$ as follows:

\vspace{-0.15in}
\[ Val_\cup(s,v) \leq Val_i(s,v) \]

Therefore we conclude the following:
\[ Val_\cup(s,v) \leq Val_i(s,v) \leq Val_\cap(s,v) \]

Note that the above result holds even when vertex $v$ is not reachable from $s$ in $G_i$ or $G_\cap$ (i.e., shortest path value is $\infty$).

Finally, in an evolving graph, an edge between a pair of nodes may be added and deleted a number of times. This type of edge will not be present in $G_{\cap}$ since it is not present in all the snapshots. However, it will be present in $G_{\cup}$. The weight of this edge can be set to the minimum of all weights encountered for this edge to obtain a safe lowerbound. 

\begin{table}[!t]
\vspace{0.05in}
\caption{Upper and Lower bounds for different algorithms.}
\label{UpperboundLowerbound}
\vspace{-0.1in}
\small
\centering
\begin{tabular}{|l|c|} \hline

Alg.
& Upperbound and Lowerbound for $Val_i(s,v)$ \\ \hline \hline

BFS
& $Val_\cup(s,v) \leq Val_i(s,v) \leq Val_\cap(s,v)$ \\ \hline

SSWP
& $Val_\cap(s,v) \leq Val_i(s,v) \leq Val_\cup(s,v)$ \\ \hline

SSNP
& $Val_\cup(s,v) \leq Val_i(s,v) \leq Val_\cap(s,v)$ \\ \hline

SSSP
& $Val_\cup(s,v) \leq Val_i(s,v) \leq Val_\cap(s,v)$ \\ \hline
 
Viterbi
& $Val_\cap(s,v) \leq Val_i(s,v) \leq Val_\cup(s,v)$ \\ \hline

\end{tabular}
\vspace{-0.15in}
\end{table}

The results of computing the lower and upper bounds for our example using the intersection and union graphs are shown in Figure~\ref{alg2}. 

In our discussion we presented the upper and lower bounds for $Val_i(s, v)$ for the SSSP algorithm. In Table~\ref{UpperboundLowerbound}, we present the bounds for various benchmarks that we use in our evaluation (see Table~\ref{benchmarks}). The upper and lower bounds for the SSSP, SSNP, and BFS algorithms are similar because we take the minimum of all possible results for a node -- the intersection graph gives us the upperbound and the union graph the lowerbound. On the other hand, for the SSWP and Viterbi algorithms, since we take the maximum of all possible results, the union graph gives us the upperbound, and the intersection graph gives us the lowerbound.

\paragraph{Step 2: Identifying UVVs -- Vertices Whose Values Remain the Same Across All Snapshots} So far we have observed that by solving a shortest path query on both $G_{\cup}$ and $G_{\cap}$ we can bound the path lengths for any vertex across all snapshots. This is illustrated in the example shown in Figure~\ref{alg2}. It is interesting to note that for many vertices, the lowerbound and upperbound match precisely. This implies that the \textbf{shortest path lengths for these vertices remain unchanged across all snapshots}. Consequently, we already have their results, and now we need to perform incremental computations to update the results of only a subset of vertices in each snapshot.

\vspace{0.05in}
\textbf{Theorem 2.} Given a shortest path query with source vertex $s$ and some vertex $v$:

\vspace{-0.15in}
\[ if \;\; Val_\cup(s,v) = Val_\cap(s,v) \;\; then\]

\vspace{-0.25in}
\[ \forall i \;[0\ldots n] \;\; Val_i(s,v) = Val_\cup(s,v) = Val_\cap(s,v), \]

\noindent
i.e., the query result value for vertex $v$ is the same for all snapshots and equal to $Val_\cup(s,v)$ (or $Val_\cap(s,v)$).

\vspace{0.1in}
\textbf{Proof.} Since $Val_\cup(s,v) = Val_\cap(s,v)$, the shortest path from $s$ to $v$ of the same length is present both in $G_\cap$ and $G_\cup$. Moreover, the presence of shortest path in $G_\cap$ implies that it is also present in all the snapshots because $G_\cap$ is the subgraph of each snapshot.

Furthermore, there cannot be any path from $s$ to $v$ in any snapshot $G_i$ that is of a shorter path length than $Val_\cap(s,v)$ even though $G_i$ contains edges that are not present in $G_\cap$. This is because if such a path existed, it would also be present in $G_\cup$ and that would contradict the fact that $Val_\cup(s,v) = Val_\cap(s,v)$.

\begin{figure*}[!t]
    \centering
    \includegraphics[width=0.62\linewidth]{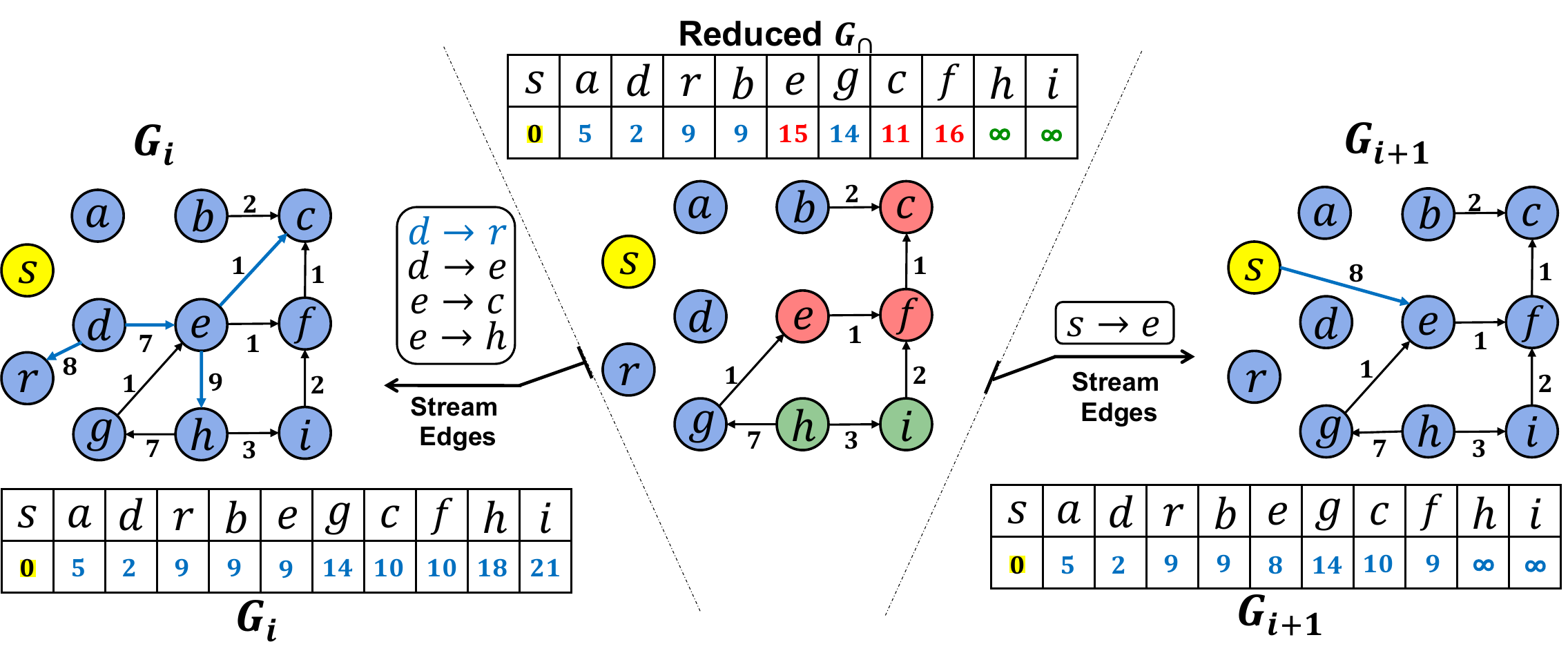}
    \vspace{-0.15in}
    \caption{Carrying out incremental computations via edge additions to obtain final query results for each snapshot. Note that for vertex $c$, the final query result is the same (10) for both snapshots while its lower and upper bounds were 9 and 11.}
    \label{alg4}
    \vspace{-0.15in}
\end{figure*}

\vspace{0.05in}
Note that when $Val_\cup(s,v) = Val_\cap(s,v)$ we have already found the shortest path value from $s$ to $v$ for all snapshots. However, when $Val_\cup(s,v) \neq Val_\cap(s,v)$, it does not imply that $Val_i(s,v)$ cannot be the same for all snapshots. 

Our algorithm is safe but not complete, that is, for a given query it does not identify all vertices for which the shortest path value remains the same across all snapshots.
In Figure~\ref{alg2} we could not identify that its value remains unchanged. This is because neither $G_{\cap}$ nor $G_{\cup}$ provide the value 10, rather they provided values 11 and 9. Yet, in Figure~\ref{alg4}, note that the shortest path value for vertex $c$ is 10 in both snapshots.

\vspace{-0.05in}
\paragraph{Step 3: Deriving Q-Relevant Subgraph} Before performing incremental computations, we can substantially reduce the size of the Intersection Graph as follows. 

\vspace{0.075in}
\begin{tabular} {|p{7.5cm}|} \hline
\textsf{For each vertex whose property value has already
been determined, that is, its lowerbound and upperbound are found to be equal, the set of its incoming edges can be removed from the graph.} \\ \hline
\end{tabular}

\vspace{0.1in}
\noindent
This is because for the vertices whose results are already known, no vertex updates are needed and the incoming edges are responsible for causing their updates can be safely eliminated. In Figure~\ref{alg2}, we show the resulting Q-Relevant Subgraph obtained after reducing the Intersection Graph $G_{\cap}$ by eliminating incoming edges of vertices with precise vertex values. The shortest path values of vertices that bootstrap the next incremental computation phase correspond to the shortest path values obtained by solving the query on $G_\cap$.

\vspace{-0.075in}
\paragraph{Step 4: Incremental Computations for Snapshots} Starting from the Q-Relevant Subgraph, and initial results computed from the Intersection Graph, we perform incremental computations to obtain precise results for both snapshots (see Figure~\ref{alg4}). Note when a batch of additions is used for incremental computation, the batch can also be further reduced by eliminating the edges whose sink is a node with already known precise solution. In Figure~\ref{alg4}, edge from $d$ to $r$ need not be streamed for snapshot $G_i$ because value for vertex $r$ is already precise and same in both snapshots.

In conclusion, while our algorithm is safe, it does not identify all vertices whose value remains the same across all snapshots. However, as the experimental results presented later in the paper will demonstrate, our algorithm is highly effective as it identifies nearly all the vertices whose property value remains unchanged across all snapshots.

\vspace{-0.1in}
\paragraph{Algorithm Summary} Algorithm~\ref{algo:QRG} shows how the QRS is found. The inputs to the algorithm are the Intersection Graph ($G_{\cap}$) and the Union Graph ($G_{\cup}$) of all the snapshots (line 1), the delta batches, which must be added to the Intersection Graph ($G_{\cap}$) to obtain the results for each snapshot of the graph. The number of delta batches equals the number of snapshots ($\Delta_{0}$, $\Delta_{1}$, ..., $\Delta_{n}$). Additionally, we should specify the query ($Q$) as an input to the algorithm. The output of the algorithm is the results of the query evaluation for each snapshot of the graph (line 3). Therefore, we have $n$ arrays to represent results of each snapshot, $R_{0}$, $R_{1}$, ..., $R_{n}$.

The first step of the algorithm is to compute the results of solving the query on the Intersection Graph ($G_{\cap}$) and the Union Graph ($G_{\cup}$). Therefore, we should define two result arrays, $R_{\cap}$ and $R_{\cup}$, to store the outcomes of the query evaluation on each graph. The size of these two arrays is proportional to the number of vertices in the graph, which is $m$ (lines 4-6). The \textsc{Compute} (Graph $G$, Query $Q$) function will evaluate the query $Q$ on the graph $G$ (lines 7-8). In the second step of the algorithm, we need to compare the results from the two value arrays $R_{\cap}$ and $R_{\cup}$. Therefore, we should use a for loop to iterate over each element of these arrays. If an element is the same in both $R_{\cap}$ and $R_{\cup}$, we should mark the vertex and add it to a set named $found$ (lines 9-15).

\begin{algorithm}[!t]
\small
\begin{algorithmic}[1]

\State{\textbf{Inputs}: 
Intersection Graph ($G_{\cap}$); Union Graph ($G_{\cup}$);}
\State{Addition Edges Batches ($\Delta_{0}$, $\Delta_{1}$, ..., $\Delta_{n}$); Query ($Q$).}
\State{\textbf{Output}: Results for Solving Query ($Q$) on Each Snapshot of the Graphs ($R_{0}$, $R_{1}$, ..., $R_{n}$).}

\State \textcolor{teal}{$\rhd$ Compute $Q$ on $G_{\cap}$ and $G_{\cup}$ ($m$ is number of vertices)}
\State{$R_{\cap}$[$1$, $m$]: array results for $G_{\cap}$}
\State{$R_{\cup}$[$1$, $m$]: array results for $G_{\cup}$}
\State{$R_{\cap}$[$1$, $m$] $\leftarrow$ \textsc{Compute} ($G_{\cap}$, $Q$)}
\State{$R_{\cup}$[$1$, $m$] $\leftarrow$ \textsc{Compute} ($G_{\cup}$, $Q$)}

\State \textcolor{teal}{$\rhd$ Find Values That Are Same On Both $G_{\cap}$ and $G_{\cup}$}
\State{$found$: set for storing vertices with precise values}
\ForEach{$i$ $\in$ [$1$, $m$]}
\If{$R_{\cap}$[$i$] == $R_{\cup}$[$i$]}
\State{$found$ $\leftarrow$ $insert (i)$}
\EndIf
\EndFor

\State \textcolor{teal}{$\rhd$ Reduce the size of $G_{\cap}$ and delta batches and Find $G_{QRS}$}
\ForEach{$v$ $\in$ $found$}
\State{\textsc{RemoveIncomingEdges}($G_{\cap}$, $v$)}
\State{\textsc{RemoveDeltaAdditionBatches}($v$)}
\EndFor
\State{$G_{QRS}$ $\leftarrow$ $G_{\cap}$}

\State \textcolor{teal}{$\rhd$ Add the Batches to $G_{QRS}$ and Find the Results}
\ForEach{$i$ $\in$ [$0$, $n$]}
\State{$R_{i}$ $\leftarrow$ $\textsc{Increment}$ ($G_{QRS}$, $\Delta_{i}$)}
\EndFor

\State \textcolor{teal}{$\rhd$ Function for removing the incoming edges}
\Function{RemoveIncomingEdges}{Graph $G$, Vertex $v$}
\ForEach{$x$ $\in$ $G[v]$.inNeighbors}
\State{remove $edge$ ($x$, $v$)}
\EndFor
\EndFunction

\State \textcolor{teal}{$\rhd$ Function for removing edges from delta batches}
\Function{RemoveDeltaAdditionBatches}{Vertex $v$}
\ForEach{$i$ $\in$ [$0$, $n$]}
\ForEach{$edge$ ($u$, $x$) $\in$ $\Delta_i$}
\If{$x$ == $v$}
\State{remove $edge$ ($u$, $v$)}
\EndIf
\EndFor
\EndFor
\EndFunction

\end{algorithmic}
\caption{Finding Q-Relevant Subgraph}
\label{algo:QRG}
\end{algorithm}

Next, we should reduce the size of the Intersection Graph ($G_{\cap}$) using the $found$ set. Currently, the $found$ set consists of all the vertices with the same value across all the snapshots. Therefore, we should remove the incoming edges of those vertices that are in the $found$ set using the \textsc{RemoveIncomingEdges} function. We should also remove the edges in the delta batches that have the same destination as those in the $found$ set using the \textsc{RemoveDeltaAddition} function (lines 16-21). Then, we can rename the reduced $G_{\cap}$ graph to $G_{QRS}$. Finally, we should incrementally add the reduced-size delta batches to the Q-Relevant Subgraph ($G_{QRS}$) to determine results of query $Q$ on each snapshot (lines 22-25).

The function \textsc{RemoveIncomingEdges}(Graph $G$, Vertex $v$) has two inputs: Graph $G$ and Vertex $v$. It iterates over each vertex in $G$, and if there is an edge leading to vertex $v$, it removes that edge (lines 26-31). The function \textsc{RemoveDeltaAddition}(Vertex $v$) takes a vertex as its input and iterates over all the edges in the delta batches. If it finds an edge to vertex $v$, it removes that edge (lines 32-41).

\section{Concurrent Incremental Computations}
Starting from the Q-Relevant Subgraph $G_{QRS}$, the query results for each snapshot can be found by incremental computation of the addition batch ($\Delta_i$).
One way to calculate query results for all snapshots is through a sequence of snapshot evaluations, i.e., $n$ rounds of incremental computation ($INCREMENT(G_{QRS},\Delta_i')$, where $i=0,...,n$ and $\Delta_i'$ is the reduced batch corresponding to $\Delta_i$). 
However, this approach has two issues: 1) resource under utilization and 2) data locality is not fully exploited.
First, incremental computation (especially for edge additions) is lightweight in comparison to full query evaluation, potentially leading to underutilization of machine resources. 
Second, same edges, either present in $G_{QRS}$ or $\Delta_i'$,  may be traversed multiple times across different snapshots, worsening cache locality.

Instead of evaluating snapshots one by one, we propose concurrent evaluation. An augmented graph with versioning and \textit{snapshot-oblivious} frontier are used for efficiency.

\begin{figure}[!t]
    \centering
    \includegraphics[width=0.95\linewidth]{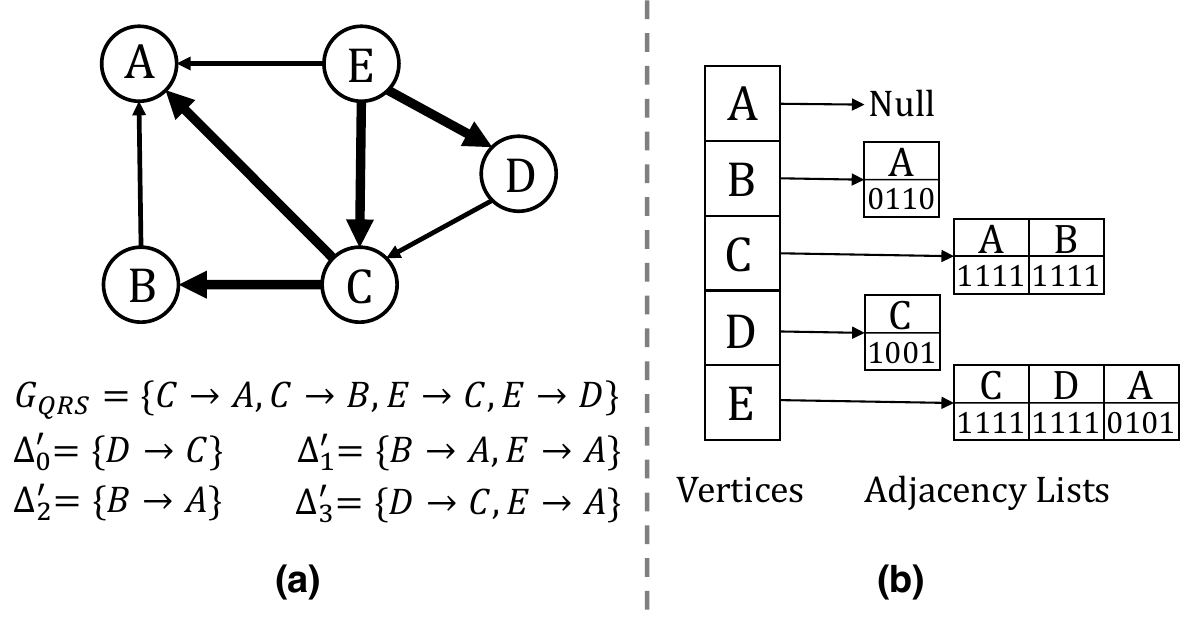}
    \vspace{-0.175in}
    \caption{(a) Versioned Graph; and (b) the augmented adjacency list (only lists for out-going edges are shown; bold edges are $G_{QRS}$ edges; and there are 4 snapshots)}
    \label{fig:vg_example}
    \vspace{-0.25in}
\end{figure}

        

\subsection{Versioned Graph Representation}
We augment the Q-Relevant Subgraph $G_{QRS}$ with extra edge versioning information to show which snapshots an edge belongs to. A 64-bit variable is used for storing such version information of an edge (more bits can be added for supporting greater than 64 snapshots). The bit at each location indicates if the edge is present in the corresponding snapshot. For edges that are common to all snapshots, their version labels are all 1s (i.e., \texttt{1111....1111}). 

The augmented versioned graph has more edges than the Q-Relevant Subgraph -- edges in $G_{QRS}$ plus the reduced addition batches ($G_{QRS}\cup \Delta_0'\cup \Delta_1'\cup ... \Delta_{n}'$). 
Since $G_{QRS}$ is reduced from the Intersection Graph ($G_{\cap}$), it contains a subset of edges in $G_{\cap}$ that are common to all snapshots. Those 
common edges are stored at the beginning of the adjacency lists, followed by snapshot-specific edges.

An  augmented graph is shown in Figure~\ref{fig:vg_example}. 
$G_{QRS}$ has four edges (edges in bold) that have \texttt{1111}s as their version value in the adjacency lists.
Four graph snapshots are embedded into the augmented graph, which can be obtained by adding $\Delta_0'$ through $\Delta_3'$ to $G_{QRS}$, respectively.
Edge $\langle D\rightarrow C\rangle$ is present in both snapshots 0 and 3, so its version number is \texttt{1001}. Edges $\langle E\rightarrow C\rangle$ and $\langle E\rightarrow D\rangle$ are common to all snapshots. 

\subsection{Concurrent Snapshot Evaluation}
Now we describe how multiple snapshots are evaluated concurrently.
The traditional graph query evaluation, the out-going edges of active vertices are evaluated by the edge function and vertices that have their values changed will be put into the frontier.
In the concurrent snapshot evaluation, there are two aspects to consider:
1) the ownership of edges that must be checked when traversing the versioned graph; and 2) the ownership of active vertices that help distinguish which vertex is active for which snapshot.
The edge ownership cannot be neglected as it affects the correctness of concurrent snapshot evaluation. We show that the ownership of active vertices can be further relaxed for better performance. In a basic concurrent evaluation design, it is intuitive to maintain a separate frontier for each snapshot since an active vertex may not be active for all snapshots; however, this introduces extra overhead due to the maintenance and access of multiple frontiers.
Instead, \texttt{UVVs} employs a design called \textit{snapshot-oblivious} frontier, inspired by recent works on concurrent graph query processing~\cite{yin2022glign,mazloumi2019multilyra}.
The \textit{snapshot-oblivious} frontier does not distinguish which vertex is active for which snapshot; given a batch of snapshots, it simply treats the vertex active for all snapshots by using a single frontier, which is the union of all separate frontiers. The correctness of \textit{snapshot-oblivious} frontier is guaranteed by the monotonic property of graph algorithms.


\begin{algorithm}[t]
\caption{Concurrent Evaluation of Snapshots}
\label{algo:concurrent-snapshot}
\begin{algorithmic}[1]
\algnotext{EndFor}
\algnotext{EndParFor}
\algnotext{EndIf}
\fontsize{8}{8.5}
\selectfont
\Function{BatchEvaluation}{$G$, $n$, $\Delta[...]$, $f$}
\State ($F_{so}$,$F_{next}$) = ($\emptyset$, $\emptyset$) \textcolor{teal}{ $\rhd$  Snapshot-oblivious frontier }
\State $R_{i\in [0:n-1]}$ = $R_{QRS}$ \textcolor{teal}{ $\rhd$  Initialize results for each snapshot  }
\ParFor{$\Delta_i$ \textbf{in} $\Delta[...]$} \textcolor{teal}{ $\rhd$  Processing addition batches } \label{line:inc_add_0}
\ParFor{$\langle u\rightarrow v, w\rangle$ \textbf{in} $\Delta_i$} 
\If{$f(\langle u\rightarrow v, w\rangle)$ improves $R_i[v]$}
    \State $R_i[v]$=$f(\langle u\rightarrow v, w\rangle)$
    \State $F_{so}$=$F_{so}\cup\{v\}$
\EndIf
\EndParFor
\EndParFor\label{line:inc_add_1}
\textcolor{teal}{$\rhd$ Concurrent snapshot evaluation }
\While {$F_{so} \neq \emptyset$}
\ParFor{$v$ \textbf{in} $F_{so}$} \label{line:snapshot_oblivious}
\For{$x$ \textbf{in} $v$'s out-going neighbors}
\For{$i$ \textbf{from} $0$ \textbf{to} $n-1$} \label{line:concurrent_eval_0}

\hspace{0.5in}\textcolor{teal}{ $\rhd$ Check if the edge belongs to snapshot $i$ }
\If{snapshotHasEdge($i$,$\langle v\rightarrow x\rangle$)}  
\If{$f(\langle v\rightarrow x, w\rangle)$ improves $R_i[x]$}
    \State $R_i[x]$=$f(\langle v\rightarrow x, w\rangle)$
    \State $F_{next}$=$F_{next}\cup\{x\}$
\EndIf
\EndIf
\EndFor
\EndFor
\EndParFor\label{line:concurrent_eval_1}
\State swap($F_{next}$, $F_{so}$)
\State $F_{next}=\emptyset$
\EndWhile
\State \Return{$R_{i\in [0:n-1]}$}
\EndFunction
\end{algorithmic}
\end{algorithm}

\begin{figure}[h]
    \centering
    \includegraphics[width=0.8\columnwidth]{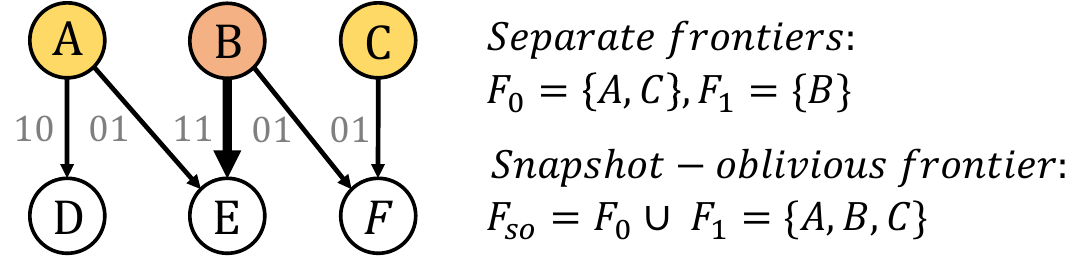}
    \vspace{-0.1in}
    \caption{\emph{Snapshot-oblivious frontier} and concurrent snapshot traversal.}
    \label{fig:sof_example}
\vspace{-0.15in}
\end{figure}

Figure~\ref{fig:sof_example} shows an example of \textit{snapshot-oblivious} frontier and concurrent snapshot traversal. Two snapshots are considered in this case with their frontiers $F_0$ and $F_1$. 
The out-going edges of active vertices in $F_0$ and $F_1$ have their ownership checked before the edge function can be applied.
For example, active vertex $A$'s out-going edge $\langle A\rightarrow E\rangle$ that is owned by snapshot $S_0$ (the version label is \texttt{01}) will be evaluated for updating the results of $S_0$, while $\langle A\rightarrow D\rangle$ will not be evaluated for $S_0$ as it only belongs to $S_1$.
Regarding the ownership of active vertices, because of the use of \textit{snapshot-oblivious} frontier $F_{so}$, vertex $B$'s out-going edges are blindly evaluated for both snapshots even if it is only active for $S_1$. 
The benefits of using \textit{snapshot-oblivious} frontier (without maintaining and checking separate frontiers) outweighs the extra computation overhead it introduces. Similar observation has been made in $\texttt{Glign}$~\cite{yin2022glign} about the \textit{query-oblivious} frontier.

The concurrent snapshot evaluation in Algorithm~\ref{algo:concurrent-snapshot} takes Q-Relevant Subgraph as input and to which addition batch $\Delta_i$ is applied to incrementally compute query results on snapshot $i$. The incremental computation for edge additions adds vertex $v$ to the frontier if a new edge $\langle u,v\rangle$ in addition batch improves the vertex value of $v$ (Line~\ref{line:inc_add_0} to ~\ref{line:inc_add_1}), e.g., a shorter shortest path is found after adding the edge.
Iterative computation continues till the frontier becomes empty. At Line~\ref{line:snapshot_oblivious}, the \textit{snapshot-oblivious} frontier does not distinguish between snapshots; it evaluates the vertex for all snapshots which may introduce redundant computation but still outweighs the overhead of tracking separate frontiers for each snapshot. Out-going edges of active vertices are evaluated for snapshots that contain the edges (Line~\ref{line:concurrent_eval_0} to ~\ref{line:concurrent_eval_1}). 


\section{System}
Based on the proposed UVV finding algorithm and concurrent snapshots query evaluation engine, we developed a system for shared-memory. To our knowledge, this is the first system of its kind that supports concurrent evolving-graph query evaluation while using a very small portion of a large graph. \textbf{\emph{Our system is built upon Risgraph as it provides the most optimized implementation of KickStarter-based incremental approach that serves as the baseline in our evaluation.}} Our UVV system consists of several components described below:

\begin{itemize}
    \item An \emph{evolving graph engine} maintains data structures of the multiple snapshots. It is built on the Risgraph ~\cite{risgraph} streaming graph engine and adopts version management of Fig~\ref{fig:vg_example}(b). Risgraph leverages a fast addition query function and uses a pull-push hybrid mechanism and our support concurrent snapshots traversal.
    \item A graph reduction phase is added to the system that reduces the graph size to create a \emph{query-relevant} graph.
    \item We employ a snapshots scheduler, which takes user queries as input and maximizes the reuse of snapshots. A snapshot-oblivious frontier mechanism is added to Risgraph. Users can pick the snapshots of interest.
    \item Finally, a simple programming interface is provided where the user can program by following the vertex-centric paradigm and providing the system with the query source and snapshots to get query results.
\end{itemize}

\textbf{System Execution Engine} Two execution models are supported: concurrent and non-concurrent. Both perform graph reduction for a given user query. As Algorithm~\ref{alg1} depicted, we query intersection graph and union graph to find UVVs. The system then deletes all incoming edges of UVV vertices to create the query-relevant subgraph. The overhead of this step is included in query evaluation time.

In the non-concurrent execution mode, the system executes a query targeting multiple snapshots in two steps: the scheduling phase and the computation phase. During the scheduling phase, the query execution plan follows Fig~\ref{fig:commongraph}, and the engine follows the scheduling order to perform addition-only incremental computation for all the queried snapshots. 

\section{Performance Evaluation}
\label{Evaluation}
We have successfully implemented our idea on top of the RisGraph~\cite{risgraph} system which is the fastest streaming system that supports incremental algorithms for handling both edge additions and deletions. We made it suitable for analyzing an evolving graph by performing incremental computations incrementally over a versioned graph representation. We conducted our experiments on a shared memory machine on Google Cloud that has two Intel Xeon 2.60 GHz processors with 48 cores and 768GB memory. All codes are compiled by g++ version 9.4.0 and we run them on Ubuntu 20.04. 

\paragraph*{Benchmarks} We evaluated five types of graph benchmarks: BFS (Breadth First Search), SSSP (Single Source Shortest Path), SSWP (Single Source Widest Path), SSNP (Single Source Narrowest Path), and Viterbi. Table~\ref{benchmarks} lists the benchmarks along with their edge functions.

\paragraph*{Graph Data Sets} We used 5 real-world input graphs, as shown in Table~\ref{graphs}. The range for the number of vertices in our input graphs is from 68M to 2.6B, and the range for the number of edges is from 4.8M to 68.3M.

\begin{table}[!t]
\caption{Benchmarks and their Edge Functions.}
\label{benchmarks}
\vspace{-0.1in}
\small
\centering
\begin{tabular}{|l|l|} \hline

Alg.
& EdgeFunction $( e(u,v) )$ \\ \hline \hline

BFS
& $CASMIN( Val(v), min( Val(u)+1, val(v) ) )$ \\ \hline

SSWP
& $CASMAX( Val(v),  min( Val(u), wt(u,v) ) )$ \\ \hline

SSNP
& $CASMIN( Val(v),  max( Val(u), wt(u,v) ) )$ \\ \hline

SSSP
& $CASMIN( Val(v),  Val(u) + wt(u,v) )$ \\ \hline

Viterbi
& $CASMAX( Val(v),  Val(u) / wt(u,v) )$ \\ \hline

\end{tabular}
\vspace{0.05in}

\captionsetup{justification=centering}
\caption{Edges and Vertices of the Input Graphs.}
\label{graphs}
\vspace{-0.1in}
\small
\centering
{\renewcommand{\arraystretch}{1.1}
\begin{tabular}{|l|c|c|c|} \hline
Input Graph & | Edges | &\!\!| Vertices |\!\! & \!\!|Avg degree|\!\!\\ \hline \hline
LiveJournal (LJ)~\cite{LJ} & 68M & 4.8M & 28.26\\ \hline
Orkut (OR)~\cite{OR} & 117M & 3.1M & 76.28\\ \hline
WikipediaLinks (Wen)~\cite{Wen}\!\!& 437M & 13.5M & 64.32\\ \hline
Twitter (TW)~\cite{TW} & 1.46B & 41.6M & 70.51\\ \hline
Friendster (Fr)~\cite{friendster} & 2.58B & 68.3M & 55\\ \hline

\end{tabular}
}
\vspace{-0.15in}
\end{table}

\begin{table}[!t]
\caption{Average Execution Time for KickStarter-based (KS) method in milliseconds, and speedup for CommonGraph (CG), Q-Relevant Subgraph (QRS), Concurrent QRS (CQRS) given 64 Snapshots and 150,000 batch size.}

\label{tab:time}
\vspace{-0.1in}
\small
\centering
{\renewcommand{\arraystretch}{1.2}
\begin{tabular}{|l|l||c|c|c|c|c|} \hline
\multicolumn{1}{|c|}{\textbf{\textsf{G}}} & \textbf{\textsf{Alg.}} & \textsf{\textbf{BFS}} & \textsf{\textbf{SSSP}} & \textsf{\textbf{SSWP}} & \textsf{\textbf{SSNP}} 
& \textsf{\textbf{Viterbi}} 
\\ \hline \hline

\multirow{4}{*}{\textsf{\textbf{LJ}}} &
    \textsc{KS} & 1562.1 & 2292.2 & 2249.6 & 2183.0 & 2677.5\\ \cline{2-7}
    & \textsc{CG} & 1.17$\times$ & 1.24$\times$ & 1.32$\times$ & 1.39$\times$ & 1.07$\times$\\ \cline{2-7}
    & \textsc{QRS} & 1.30$\times$ & 1.53$\times$ & 2.08$\times$ & 2.43$\times$ & 1.67$\times$\\ \cline{2-7}
    & \textsc{CQRS} & 3.60$\times$ & 2.01$\times$ & 8.09$\times$ & 7.62$\times$ & 8.87$\times$
    \\ \hline \hline
\multirow{4}{*}{\textsf{\textbf{OR}}} &
     \textsc{KS} & 905.9 & 1361.7 & 1499.1 & 1330.5 & 1310.2\\ \cline{2-7}
    & \textsc{CG} & 1.11$\times$ & 1.06$\times$ & 1.44$\times$ & 1.17$\times$ & 1.08$\times$\\ \cline{2-7}
    & \textsc{QRS} & 1.56$\times$ & 1.14$\times$ & 1.60$\times$ & 1.53$\times$ & 1.25$\times$\\ \cline{2-7}
    & \textsc{CQRS} & 3.69$\times$ & 3.72$\times$ & 8.37$\times$ & 3.76$\times$ & 3.03$\times$
    \\ \hline \hline
\multirow{4}{*}{\textsf{\textbf{Wen}}} &
     \textsc{KS} & 919.3 & 1850.1 & 1391.9 & 994.2 & 1007.3\\ \cline{2-7}
    & \textsc{CG} & 1.23$\times$ & 1.42$\times$ & 1.37$\times$ & 1.46$\times$ & 1.10$\times$\\ \cline{2-7}
    & \textsc{QRS} & 1.55$\times$ & 1.77$\times$ & 1.85$\times$ & 2.25$\times$ & 1.52$\times$\\ \cline{2-7}
    & \textsc{CQRS} & 3.07$\times$ & 6.19$\times$ & 11.7$\times$ & 5.90$\times$ & 3.56$\times$
    \\ \hline \hline
\multirow{4}{*}{\textsf{\textbf{TW}}} &
      \textsc{KS} & 971.8 & 1997.6 & 1993.2 & 1304.9 & 1774.7\\ \cline{2-7}
    & \textsc{CG} & 1.06$\times$ & 1.86$\times$ & 1.61$\times$ & 1.20$\times$ & 1.03$\times$\\ \cline{2-7}
    & \textsc{QRS} & 1.44$\times$ & 2.94$\times$ & 3.44$\times$ & 1.51$\times$ & 2.61$\times$\\ \cline{2-7}
    & \textsc{CQRS} & 3.77$\times$ & 8.88$\times$ & 10.23$\times$ & 6.01$\times$ & 8.25$\times$
    \\ \hline \hline
\multirow{4}{*}{\textsf{\textbf{Fr}}} &
      \textsc{KS} & 1349.8 & 2680.9 & 1951.6 & 1824.5 & 2968.5\\ \cline{2-7}
    & \textsc{CG} & 1.32$\times$ & 1.13$\times$ & 1.25$\times$ & 1.2$\times$ & 1.16$\times$\\ \cline{2-7}
    & \textsc{QRS} & 2.08$\times$ & 1.4$\times$ & 3.57$\times$ & 4.3$\times$ & 1.82$\times$\\ \cline{2-7}
    & \textsc{CQRS} & 6.5$\times$ & 8.19$\times$ & 9.57$\times$ & 12.23$\times$ & 11.95$\times$
    \\ \hline
\end{tabular}
}
\vspace{-0.3in}
\end{table}

\vspace{0.05in}
\subsection{Speedups}
    In our experiments we compare the execution times of query evaluation for five input graphs and 5 benchmarks across a sequence of 64 snapshots. Between two consecutive snapshots there are 150,000 edge updates, half of them are additions and half are deletions. 
We present our results in Table~\ref{tab:time}. 

\textbf{KickStarter-based (KS)} baseline incremental approach does full computation on the first snapshot followed by repeated incremental processing of addition and deletion batches to get results for subsequent snapshots (see Figure~\ref{fig:evolving_approaches}(b)). The first row of Table~\ref{tab:time} for each benchmark displays the KS execution times in milliseconds. Our implementation is based on Risgraph as it represents the most optimized implementation of KickStarter.

For \textbf{CommonGraph (CG)} we present results for the best performing implementation~\cite{CommonGraph} which is the work-sharing approach and present its speedups over KS in the second row of Table~\ref{tab:time} for each benchmark. As shown in Table~\ref{tab:time}, we did not observe a substantial speedups when implementing from CommonGraph because the RisGraph system is highly optimized, and the deletion operation is not as expensive compared to prior systems. The third row of Table~\ref{tab:time} gives the speedup achieved by the \textbf{Q-Relevant Subgraph work-sharing (QRS)} method over the KickStarter-based method. The QRS method significantly reduces the size of the graph over which incremental computations are performed and hence leads to a considerable speedup in obtaining results for individual snapshots. \textbf{\emph{However, there is overhead associated with generating the QRS graph (overhead cost) which we take into account by including it in the total query evaluation time. }}\textbf{The QRS approach yields speedups ranging from 1.25$\times$ to 4.3$\times$ over KS.}

The fourth row of Table~\ref{tab:time} gives the speed-up achieved by \textbf{Concurrent Q-Relevant Subgraph (CQRS)} over the KickStarter-based implementation. In this technique, we concurrently add the additional delta batches to the QRG. \textbf{\emph{We have included the time for QRS generation (overhead cost) in query evaluation time.}} \textbf{CQRS yields speedups ranging from 2.01$\times$ and 12.23$\times$ over KS.}

\begin{figure*}[!t]
\vspace{0.05in}

    \centering
    \includegraphics[width=0.99\linewidth]{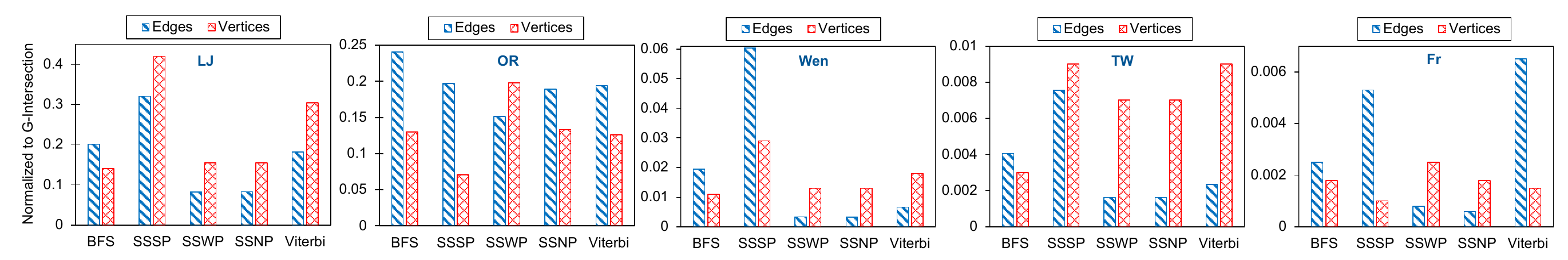}
    \vspace{-0.15in}
    \caption{Remaining fraction of edges in QRS (blue bars) and fraction of vertices whose values are computed via incremental computation (red bars), both normalized with respect to edges and vertices in $G_{\cap}$. This experiment uses 64 snapshots and a batch size of 150,000 edge updates (half additions and half deletions).}
    \label{reduction}

    \centering
    \includegraphics[width=0.99\linewidth]{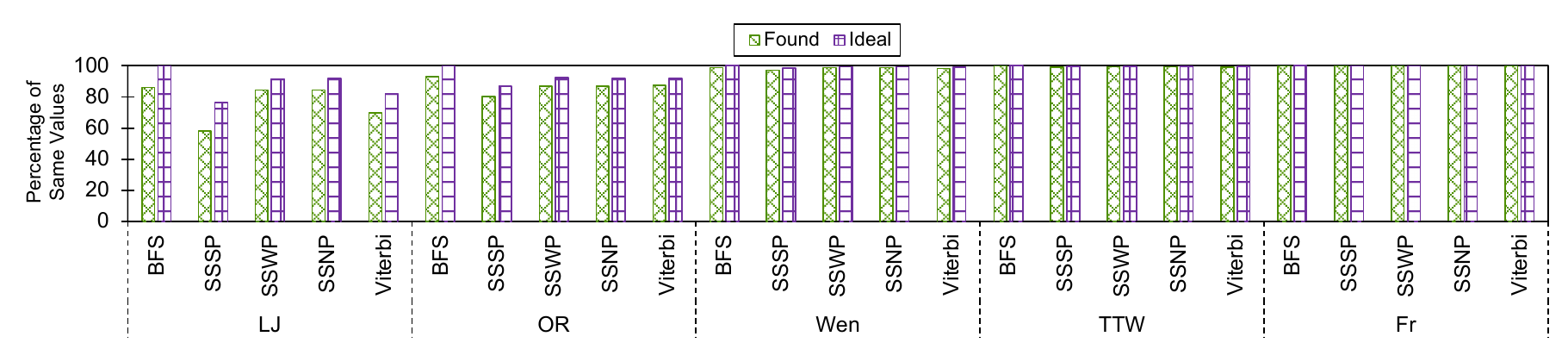}
    \vspace{-0.15in}
    \caption{Percentage of vertices that are UVVs (purple bars) vs. percentage of UVVs successfully detected by our algorithm (green bars). This experiment uses 64 snapshots and a batch size of 150,000 edge updates (half additions and half deletions).}
    \label{accuracy}

    \centering  
    \includegraphics[width=0.99\linewidth]{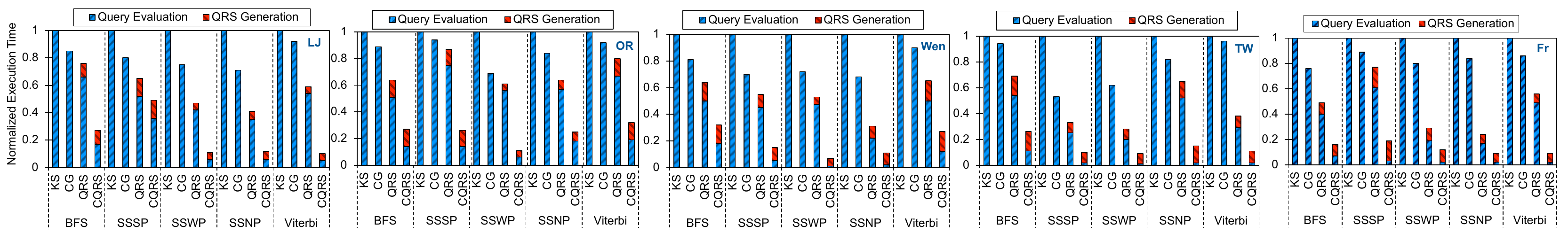}
    \vspace{-0.1in}
    \caption{The normalized execution times for Kickstarter-based (KS), CommonGraph Work-Sharing (CG), Q-Relevant Subgraph (QRS), and Concurrent Q-Relevant Subgraph (CQRS) are presented with a breakdown of QRS generation time (overhead cost) and query evaluation times for 64 snapshots and 150,000 edge updates (half additions and half deletions).}
    \label{overhead}
    \vspace{-0.15in}  
\end{figure*}

\subsection{UVV Detection and QRS Generation}
After detecting UVVs, we are able to determine the fraction of vertices who results need to be updated incrementally. After removing incoming edges of UVV vertices, the fraction of edges remaining in QRS is also known. The percentage of vertices ranges from 0.3\% to 42\% and edges involved incremental computation ranges from 0.16\% to 32\% (see Figure~\ref{reduction}). This large reduction is due to high accuracy with which UVVs are detected (see Figure~\ref{accuracy}).

Next we present the portion of QRS and CQRS execution times spent on QRS generation. Figure~\ref{overhead} shows the execution times for all methods normalized to the KickStarter-based approach (KS). The red segments in the QRS and CQRS bars represent the QRS generation times. 
The QRS generation consists of four steps. First, we solve the query on the intersection graph. Then, we incrementally add the missing edges to the intersection graph to obtain the results on the union graph. We compare the two value arrays, and if we find the same value for a node in both the $G_{\cup}$ and $G_{\cap}$ arrays, we should remove the incoming edges for that vertex. However, due to the much larger number of matches than mismatches, we implement this step in reverse. Instead of removing the incoming edges of the matching values, we add incoming edges for the mismatching values. On average, the QRS generation time accounts for 18.45\%/56.01\% of the total execution time for QRS/CQRS.


\subsection{Sensitivity to Number of Snapshots}
\begin{figure}[!ht]
    \centering
    \includegraphics[width=0.925\linewidth]{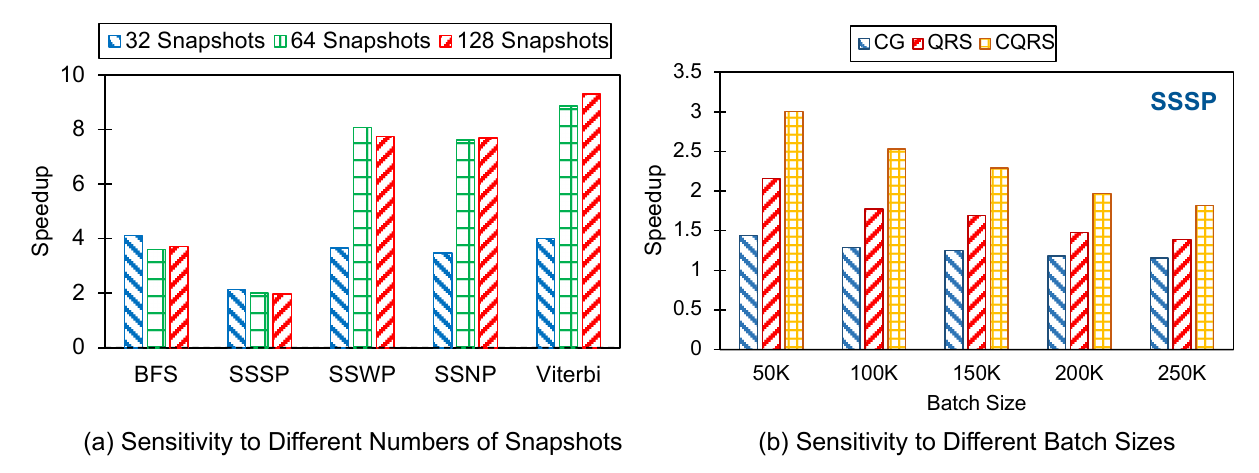}
    \vspace{-0.175in}
    \caption{(a) Sensitivity to the number of snapshots (32, 64, and 128) with batches of 150K edge updates for the LiveJournal graph; (b) Sensitivity to batch sizes of 50K, 100K, 150K, 200K, and 250K for the LiveJournal and the SSSP algorithm.}
    \label{sensitivity-snapshots}
    \vspace{-0.25in}
\end{figure}

We studied the performance of UVV for 32, 64, and 128 snapshots. Figure~\ref{sensitivity-snapshots}(a) shows high speedups across varying number of snapshots, benchmarks, and graphs. Speedups for 64 and 128 snapshots are close. While the speedups for 32 snapshots are lower. This is because the resources available on our server are not fully utilized by 32 snapshots but are by 64 and 128 snapshots. We studied the sensitivity to different batch sizes for the LiveJournal graph. We chose LiveJournal for this experiment because it is our smallest graph, allowing us to observe the effects of changes in batch sizes more clearly. Figure~\ref{sensitivity-snapshots}(b) shows that larger batch sizes give lower speedups because they lead to more updates and fewer UVVs.

\section{Related Work}

{\bf Evolving graphs.} Recent works on dynamic graphs are Common Graph~\cite{CommonGraph}, RisGraph~\cite{risgraph}, and Tegra~\cite{tegra}. Common Graph transforms all deletions into additions by using the common subgraph present in all snapshots. Query is executed on this graph and then  missing edges are added incrementally for each snapshot. In this paper we exploited UVVs while preserving all the benefits of the Common Graph. In addition, unlike Common Graph, incremental computations are done concurrently over a versioned graph representation.

RisGraph and Tegra \emph{explicitly} process deletions and additions. For deletions they use the KickStarter~\cite{kickstarter} algorithm. RisGraph uses a new data structure for quick edge additions and removals, but this leads to memory size increase of 3.25x to 3.38x. Tegra offers an API for efficient querying over time windows, using a compact in-memory graph format. Both RisGraph and Tegra use algorithms from streaming systems to facilitate incremental computations. GraphOne~\cite{GraphOne} and Aspen ~\cite{Aspen} are other systems supporting dynamic and streaming graphs, while Chronos~\cite{chronos} and FA+PA~\cite{evog-taco} optimize memory and computation costs. But they lack edge deletion support when the goal is to perform query evaluation. LiveGraph~\cite{LiveGraph} presents an innovative approach to handling dynamic graph updates using transactionally consistent adjacency lists, significantly improving performance for evolving graphs. Other systems use graph sharing when simultaneous evaluating multiple queries on one graph version~\cite{krill, glign, graphm}.

\vspace{-0.075in}
\paragraph*{Streaming graph analytics} These algorithms keep one version of the graph that is continuously updated and the results of an query which are progressively updated as new batches of changes are made to the graph. There are two key aspects of these systems, fast graph mutation and fast incremental query evaluation. In contrast, all versions of an evolving graph (i.e., multiple snapshots) are available at the outset and thus graph mutation is not a concern as a multi-versioned graph representation is created. However, the incremental query evaluations developed for streaming graph systems are leveraged by evolving graph systems (e.g., Tegra~\cite{tegra}, Common Graph~\cite{CommonGraph}).  The primary focus of these systems is on incremental computation, specifically on how to effectively update query outcomes. Early streaming platforms like Kineograph~\cite{kineograph}, Naiad~\cite{naiad}, Tornado~\cite{tornado}, and Tripoline~\cite{tripoline} only support edge additions. However, Kickstarter~\cite{kickstarter} and GraphBolt~\cite{graphbolt} support edge deletions too.

\section{Conclusion}
We identified a new opportunity for optimizing the evaluation of an evolving graph query over large number of snapshots. We showed that a large fraction of vertices whose query results do not change can be identified and computed once for all snapshots and exploited to minimize incremental computations performed concurrently for all snapshots. 


\balance
\bibliography{refs}

\end{document}